\documentclass[fleqn,usenatbib]{mnras}
\usepackage[T1]{fontenc}
\usepackage{graphicx}	
\usepackage{amsmath}	
\usepackage{amssymb}	

\usepackage{times}
\usepackage{epsfig}
\usepackage{psfrag}
\usepackage{ulem}
\usepackage{tikz}
\usepackage{etaremune}
\usepackage{pifont}
\usepackage{colortbl}
\usepackage{soul}

\usepackage{hyperref}

\title[GWs from LGRBs jet propagation]{Gravitational Waves from the Propagation of Long Gamma-Ray Burst jets}

\author[Urrutia, De Colle, Moreno \& Zanolin]{
Gerardo Urrutia$^{1}$\thanks{E-mail: gerardo.urrutia@nucleares.unam.mx},
Fabio De Colle$^{1}$\thanks{E-mail: fabio@nucleares.unam.mx},
Claudia Moreno$^{2,3}$\thanks{E-mail: claudia.moreno@academico.udg.mx},
, and
Michele Zanolin$^{3}$
\thanks{E-mail:zanolinm@erau.edu}
\\
$^{1}$Instituto de Ciencias Nucleares, Universidad Nacional Aut{\'o}noma de M{\'e}xico, A. P. 70-543 04510 D.F. Mexico\\
$^{2}$Departamento de F\'isica,
Centro Universitario de Ciencias Exactas e Ingenier\'ias, Universidad de Guadalajara\\
Av. Revoluci\'on 1500, Colonia Ol\'impica C.P. 44430, Guadalajara, Jalisco, M\'exico\\
$^{3}$Embry-Riddle Aeronautical University, Prescott, AZ 86301, USA\\
}

\date{Accepted XXX. Received YYY; in original form ZZZ}

\pubyear{2022}

\begin{document}
\label{firstpage}
\pagerange{\pageref{firstpage}--\pageref{lastpage}}
\maketitle


\begin{abstract}
Gamma-ray bursts (GRBs) are produced during the propagation of ultra-relativistic jets. It is challenging to study the jet close to the central source, due to the high opacity of the medium. In this paper, we present numerical simulations of relativistic jets propagating through a massive, stripped envelope star associated to long GRBs, breaking out of the star and accelerating into the circumstellar medium. We compute the gravitational wave (GW) signal  resulting from the propagation of the jet through the star and the circumstellar medium. We show that key parameters of the jet propagation can be directly determined by the GW signal. The signal presents a first peak corresponding to the jet duration and a second peak which corresponds to the break-out time for an observer located close to the jet axis (which in turn depends on the stellar size), or to much larger times (corresponding to the end of the acceleration phase) for off-axis observers. We also show that the slope of the GW signal before and around the first peak tracks the jet luminosity history and the structure of the progenitor star. The amplitude of the GW signal is $h_+D \sim$ hundreds to several thousands cm. Although this signal, for extragalactic sources, is outside the range of detectability of current GW detectors, it can be detected by future instruments as BBO, DECIGO and ALIA. Our results illustrate that future detections of GW associated to GRB jets may represent a revolution in our understanding of this phenomenon.

\end{abstract}


\maketitle


\section{Introduction}


Gamma-ray bursts (GRBs) are extremely luminous pulses of gamma-rays (with an isotropic energy of 10$^{51}-10^{54}$ ergs) lasting typically from $\sim$ a fraction of a second to $\sim$ hundreds of seconds. GRBs are classified based on their duration. Short GRB (SGRBs), lasting $\lesssim 2$~s, are typically produced during the coalescence of neutron stars (NS), while long GRBs (LGRBs), lasting $\gtrsim 2$~s, are in several cases associated to the collapse of massive stars and their explosion as type Ic supernovae (SNe) (for a review, see, e.g., \citealt{Kumar_2015}). Recent observations of a kilonova associated to GRB211211a showed that the usual identification of different progenitors mainly based on the GRB duration can be misleading \citep[][]{Gao2022,Troja2022}. 

The gamma-ray emission observed in these events is produced by highly relativistic jet, moving with Lorentz factors $\Gamma_j \sim$ 100 - 1000. 
These jets are ejected from a black hole or a magnetar (the so-called ``central-engine'') formed during the collapse of a massive star (see, e.g. \citealt{HorthBloom2012,Cano-etal2017}) or as a result of the coalescence of a binary NS system (see, e.g., \citealt{Berger2014}).

Once the jet is ejected from the central engine, it propagates through the dense, optically thick surrounding medium formed by the progenitor star or the debris of the binary NS system, before breaking out at distances of $\sim 10^{10}-10^{11}$~cm. Theoretical studies show that, during this phase, the jet moves with sub-relativistic velocities ($\sim$ 0.1 - 0.5 $c$), being $c$ the light speed \citep[e.g.,][]{Bromberg2011,NakarPiran2017,Decolle2018a}.
When the jet breaks out from the dense environment, it accelerates to large jet Lorentz factors $\Gamma_j$  ($\sim E_j/M_j c^2$ where $E_j$ and $M_j$ are the jet energy and mass), before emitting the observed gamma radiation at larger distances from the central engine ($\gtrsim 10^{13}-10^{15}$~cm), once the hot plasma becomes optically thin to gamma-ray radiation.

The prompt gamma-ray emission is followed by a multi-wavelength afterglow emission covering the full electromagnetic spectrum, from radio to X-rays, and lasting from minutes to several years. Thus, the late phases of evolution of the relativistic jets (from $\sim 10^{13}$ cm to $\gtrsim 10^{18}$ cm) can be studied by analyzing these rich electromagnetic signatures (see, e.g., \citealt{Kumar_2015} and references therein). On the other hand, it is much more difficult to study the early phases of evolution of the jet, corresponding to distances $\lesssim 10^{10}-10^{11}$ cm, as the high densities make the jet plasma optically thick to electromagnetic radiation. In particular,  only neutrinos (e.g., \citealt{Kimura2022}) and GWs could probe directly the behaviour of the jet while it is crossing the dense environment. 

In addition to oscillating GWs signals associated to the coalescence of compact objects \citep{abbot2017NSmerger}, the possibility of detecting non-oscillating, low frequency signals (the so-called ``memory'' signal produced by unbound material over timescales $\gtrsim 1$ s), has been proposed long time ago \citep{1987Natur.327..123B}. 
These ``memory'' signals have been studied extensively, e.g., in the context of supernovae (SNe) explosions \citep[e.g.][]{Kotake:2005zn,Murphy2009,muller2012,Muller2013,Wongwathanarat2015,Yakunin2015,Powell2019,Hubner2020,Mezzacappa2020,Richardson_2022}. 

The focus of these studies was to discuss under which circumstances (in terms of specific instrument and signal morphology) the memory component of the signal spectral density is above the interferometric noise spectral density \citep[see, e.g.,][]{Moore2014}. This is a semiquantitative measure of the detectability of the memory (in the sense that it is an important metric but it is not related to a specific alghorithm). It is also worth stressing that for detectability the whole spectrum of the memory development over time matters, not just the zero frequency component produced by the asymptotic value.

Previous studies of the GWs produced by GRB jets have focused on the propagation of the jet through the dense envelope, or to the acceleration of the jet after the break-out \citep{Segalis_2001,Sago_2004,Sun2012,Akiba2013,piran13,Du_2018,Yu_2020, piran21}.
These studies have shown that the amplitude of the GW increases with time due to the continuous injection of energy into the jet from the central engine, or due to the jet acceleration once it expands through the environment.

Previous studies \citep{Segalis_2001,Sago_2004,Sun2012,Akiba2013,piran13,Du_2018,Yu_2020,piran21} estimating the GW memory from GRB jets were based on simple analytical and/or semi-analytical estimations. Although these calculations provide a qualitative understanding of the GW memory, quantitative estimations can be obtained only by detailed numerical calculations. 

In this work, we study the propagation of relativistic jets associated to LGRBs through the progenitor star, and its propagation through the wind of the progenitor star up to large distances ($10^{13}$~cm). We compute the resulting GW signal as a function of time and observer angle (with respect to the main axis of the jet). We also consider the possible presence of a supernova component, and how its GW signal is affected by the presence of the jet. As we will discuss below, although the simulations presented refer to the LGRB case (in which the jet is propagating through a massive progenitor star), the expected GW signal will be qualitatively similar in short GRBs. 

The paper is structured as follows: in Section \ref{sec:methos} we discuss the initial conditions of the hydrodynamic simulations, and the methods used to compute the GW directly from the simulations. Section \ref{sec:results} presents the results of the calculations, in particular, the jet dynamics as the jets propagate through the progenitor and its environment, and the calculation of the resulting GW. In section \ref{sec:discussion} we discuss our results, in the context of present and future GW detectors. Our conclusions are presented in section \ref{sec:conclusions}.

\section{Methods}\label{sec:methos}

\subsection{Numerical simulations}

\begin{table}
  \centering
    \begin{tabular}{|c|c|c|c|}
        \hline
        Scenario & $t_{\rm inj}$ (s) & Energy (erg) & Progenitor  \\ \hline \hline
        Successful Jet 1 & 10 & $10^{51}\,$ & 12TH  \\
        Successful Jet 2 & 2.5 & $10^{52}\,$& 16TH \\        
        Failed Jet & 10 & $10^{51}\,$ & 12TH\\
        Supernova & 1 & $10^{52}$ & 12TH\\
        Jet + Supernova & 10 & $10^{51}$ & 12TH \\
\hline
    \end{tabular}
    \caption{Numerical simulations presented in this paper. The columns refer to: the scenarios considered, the time during which the jet/SN is injected into the computational box, its energy, and the progenitor star (see the main text for a detailed description of each model). The progenitors 12TH and 16TH correspond to  12~M$_{\odot}$ and 16~M$_{\odot}$ initial masses, respectively.}
    \label{tab:models}
\end{table}

We study the first 300 s of evolution of  relativistic GRB jets, associated with massive stellar collapse, by running a series of numerical simulations. The simulations employ the adaptive mesh refinement code {\it Mezcal} \citep[]{decolle12}, which integrates the special relativistic, hydrodynamics equations by using a second-order (both in space and time), shock-capturing scheme.

We consider five scenarios (summarised in Table \ref{tab:models}): an asymmetric supernova (the ``supernova'' model), two successful jets without a SN associated (the ``successful jet 1'' and ``successful jet 2'' models), differing by their duration and total energy, a successful jet associated to a SN (the ``jet + supernova'' model), and a failed jet not associated to a SN (the ``failed jet'' model).

The numerical simulations (see Table \ref{tab:models}) employ two dimensional (2D), cylindrical (axisymmetric) coordinates. In all the models, the computational box extends from $(r,z) = 0$~cm to ($r_{\rm max},z_{\rm max})= 10^{13}\,$~cm, and is resolved by employing $40 \times 40$ cells at the coarsest level of refinement and $17$ levels of refinement, corresponding to a maximum resolution of $\Delta r_{\rm min}=\Delta z_{\rm min} = 3.8 \times 10^6\,$cm. We set the density in the computational box by considering the pre-collapse stellar models 12TH and 16TH taken from \citet[]{WoosleyHeger_2006}. These models\footnote{Long GRBs are associated to broad-line, type Ic SNe, which are produced during the collapse of massive, compact Wolf-Rayet stars.} corresponds to stripped-envelope progenitor stars with stellar masses $M_\star=$ 9.23 $M_\odot$ and  11.45 $M_\odot$ and stellar radii  $R_\star=4.5\times 10^{10}$~cm and  $9\times 10^{10}$~cm for the 12TH and the 16TH models respectively. For radial distances $r > R_{\star}$, we consider a medium shaped by the wind of the Wolf-Rayet progenitor, i.e. with a density
\begin{equation}
    \rho(r) = \frac{\dot{M}_w}{4\pi r^2 v_w}, 
\end{equation}
being  $\dot{M}_w=10^{-5}\,$ M$_\odot$ yr$^{-1}$ and $v_{\rm w}=10^3\,$ km s$^{-1}$ typical values for the mass-loss rate and the velocity of the wind from a Wolf-Rayet star \citep[e.g.,][]{Vink2011}.
The pressure in both the star and the wind is negligible (as in strong shock it does not affect the shock dynamics) and it is set as $p=10^{-5} \rho c^2$.

In all except the ``supernova'' model, the relativistic jet is injected from an inner boundary located at $r_{\rm in}=5 \times 10^8\,$cm, with a jet Lorentz factor $\Gamma_{j}=$10.
The jet energy is largely dominated by thermal energy, with the jet pressure given as,
\begin{equation}
    p_j= \frac{\rho_j c^2}{4} \left( \frac{\Gamma_{\infty}}{\Gamma_j} -1 \right), 
\end{equation}
being $\rho_j$ the jet mass density and $\Gamma_{\infty}=100$ the asymptotic jet velocity, eventually achieved once the jet breaks out of the star and accelerates by converting its thermal to kinetic energy. In two of the simulations (differing by the presence of a SN and indicated in Table \ref{tab:models} as ``successful jet 1'' and ``jet + supernova''), we inject the jet during $t_j=10$~s, such that its total energy is $E_j = 10^{51}$~erg and its luminosity is $L_j=10^{50}$~erg~s$^{-1}$, while in one model (the ``successful jet 2'' model) we inject the jet during $t_j=2.5$~s with a total energy of $E_j = 10^{52}$~erg, corresponding to a much larger luminosity $L_j=4\times10^{51}$~erg~s$^{-1}$. In all these cases the {\rm jet opening angle} is $\theta_j=0.1\,$rad and, as we will discuss in detail below, the jet successfully breaks out of the star and accelerates to highly relativistic speeds through the progenitor wind. We also consider a simulation in which the jet also lasts for $t_j=10$~s, with a total energy $E_j = 10^{51}$~erg, but with a larger jet opening angle $\theta_j=0.2\,$rad (the ``failed jet'' model). In this case, the jet will not be able to break out successfully from the star. We refer to this case as the choked or failed GRB case.

To study how the GW memory signal is affected by the presence of both a SN and a GRB, we also inject, in two of the five simulations (``supernova'' and ``jet + supernova'' models, see table \ref{tab:models}), a supernova shock front from the same inner boundary at $t=0$~s.
Following \citet[]{DeColle2021} and \citet{Urrutia2022a}, we inject, from $r_{\rm in}$, a SN shock front during $t_{\rm sn}=0.1$~s, with a total energy of $E_{\rm sn}=4 \times 10^{51}$~erg and a mass $M_{\rm sn}=0.1M_{\odot}$. We assume that 10\% of the SN energy is thermal, while 90\% is kinetic. Type Ic, broad-line SNe associated to long GRBs present a certain degree of asymmetry (as inferred from polarization measurements,  see, e.g., \citealt{Maund2007,Tanaka_2017}, or by the analysis of line emission during the nebular phase, see, e.g., \citealt{Taubenberger2009}). To qualitatively reproduce this asymmetry, we set an angular dependence for the energy injected in the SN as $E_{\rm SN}(\theta) \propto \cos^2 \theta$, being $\theta$ the polar angle measured with respect to the $z$-axis. 

In the ``jet + supernova'' model, in which both SN and jet are present, the jet is injected with a delay of 1 s with respect to the SN. The origin of the SN associated to GRBs is debated. The models proposed include a wind from a collapsar disk \citep{MacFadyenWoosley_1999}, energy ejection from a magnetar \citep[e.g.,][]{Metzger2015}, or the jittering jet mechanism \citep[e.g.,][]{papishSoker2014}; see also the discussion by \citet[]{DeColle2021}. Thus, the time delay between the SN and the jet is uncertain.

\subsection{Gravitational wave signals}\label{sec:2.2} 

\begin{figure}
    \centering
   \includegraphics[scale=0.32]{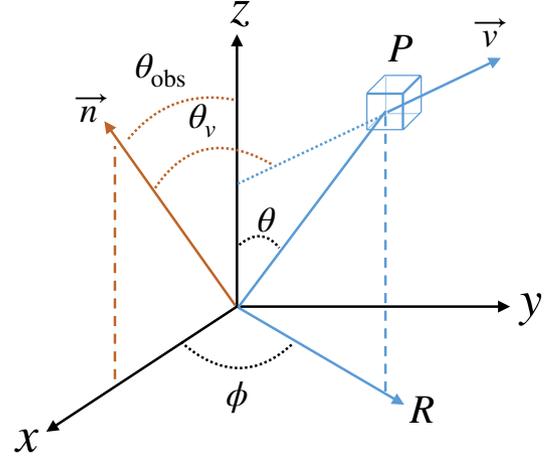}
    \caption{Schematic representation of the geometry of the problem. A fluid element $P$, located at angles $\phi, \theta$ with respect to the $x$- and $z$-axis respectively, is moving with a velocity $\vec{v}$. The observer is located along the direction $\hat{n}$, in the plane $xz$ and forming an angle $\theta_{\rm obs}$ with respect to the $z$-axis. The directions of the observer $\hat{n}$ and of the velocity vector $\vec{v}$ are separated by an angle $\theta_v$, i.e. $\cos \theta_v = \hat{n}\cdot \hat{v}$. The simulations presented in this paper are computed in two-dimensional, axisymmetric cylindrical coordinates (the $Rz$ plane shown in the figure), so that the three dimensional structure is reconstructed by rotating along the $\phi$ direction the snapshots of the numerical simulations.} 
    \label{fig1}
\end{figure}

We consider a system of reference centered on the central engine, being the $z$ axis the main axis of propagation of the jet (see Figure \ref{fig1}). The direction of the observer is defined by the unit vector $\hat{n}=(\sin \theta_{\rm obs}, 0, \cos \theta_{\rm obs})$, where $\theta_{\rm obs}$ is the angle between the direction of the observer and the $z$-axis. We rotate the $x$ and $y$ axis such that $\hat{n}$ is located in the $x,z$ plane. Thus, the axes $\hat{n}$, $y$ and $x'$ (rotated by an angle $\theta_{\rm obs}$ with respect to $x$) define a system of reference in the observer frame.
We consider a fluid element $P$, at the position $\hat{r}=(\sin\theta \cos \phi, \sin\theta \sin \phi, \cos\theta)$, moving with a velocity $\vec{v}=(v_R \cos\phi, v_R \sin\phi, v_z)$, where  $v_R,v_z$ are the fluid velocities along the radial and vertical axis of the cylindrical system of reference (see Figure \ref{fig1}). While in previous studies the velocity of the fluid element has been fixed as vertical of radial, in this paper we leave it completely general, and determined directly from the numerical simulations. 

\citealt{1987Natur.327..123B,Segalis_2001} obtained explicit expressions for the GW memory polarization components $h_+$ and $h_\times$ in the transverse-traceless (TT) gauge. The explicit expressions for $h_+$ and $h_\times$ are:
\begin{eqnarray}
    h_+\equiv h_{xx}^{TT}=-h_{yy}^{TT}&=&\frac{2G}{c^4}\frac{E}{D} \frac{\beta^2\sin^2\theta_v}{1-\beta \cos \theta_v}\cos 2 \Phi \;, 
    \label{eqn:new_h+}
    \\
    h_\times \equiv h_{xy}^{TT}=h_{yx}^{TT}&=&\frac{2G}{c^4}\frac{E}{D} \frac{\beta^2\sin^2\theta_v}{1-\beta \cos \theta_v}\sin 2 \Phi \;,
    \label{eqn:new_hx}
\end{eqnarray}
where $G$ is the gravitational constant, $D$ the distance between the object and the observer, $\beta=v/c$ is the velocity normalized with respect to the speed of light, $\theta_v$ is the angle between the direction of the observer and the direction of the velocity vector, i.e.
\begin{equation}
    \cos\theta_v = \hat{n}\cdot \hat{\beta} = (\beta_R \sin \theta_{\rm obs} \cos\phi + \beta_z \cos \theta_{\rm obs})/\beta \;,
    \label{eq:nv}
\end{equation}
$E = (\rho H \gamma^2 c^2-p)\Delta V$ is the energy of the fluid element, being $\rho$ the mass density, $\gamma$ the Lorentz factor, $p$ the pressure, $H=1+4p/(\rho c^2)$ the specific enthalpy (by considering a hot plasma with an adiabatic index $\Gamma_{\rm ad}=4/3$), $\Delta V$ the volume of the fluid element which induces the metric perturbation, and $\Phi$ is the polar coordinate, measured in the observer frame.  

To find the value of $\Phi$, we  consider the following geometric relations between the angles evaluated in the observer frames (indicating the azimuthal and polar directions by the capital Greek letters $\Phi$ and $\Theta$ respectively) and those in the laboratory frame (e.g., the frame centered on the central engine; see,  \citealt{Akiba2013}):
\begin{eqnarray}
 \cos{\Theta} = \hat{n}\cdot\hat{r} &=& \sin\theta\cos \phi\sin\theta_{\rm obs} +  \cos\theta\cos\theta_{\rm obs}, \\
 \sin\theta \sin\phi &=& \sin\Theta \sin\Phi, \\
\sin\theta \cos\phi &=&  \sin\Theta \cos\Phi\cos\theta_{\rm obs} + \cos\Theta \sin\theta_{\rm obs}\:,
\end{eqnarray}
which lead to 
\begin{eqnarray}
 \sin(2\Phi) = \nonumber  \\  2 \sin\theta
 \sin\phi \left(\frac{
  \sin\theta\cos \phi \cos\theta_{\rm obs} - \cos\theta\sin\theta_{\rm obs}
 }{\sin^2\Theta}\right),
 \label{eqn:angle_rel_1}
 \\
 \cos(2\Phi) = \nonumber\\ \frac{
 (
 \sin\theta \cos\phi\cos\theta_{\rm obs} - \cos\theta\sin\theta_{\rm obs} )^2 -\sin^2\theta \sin^2\phi }{\sin^2\Theta} .
\label{eqn:angle_rel_2}
\end{eqnarray}

\begin{figure*}
    \centering
    \includegraphics[scale=0.29]{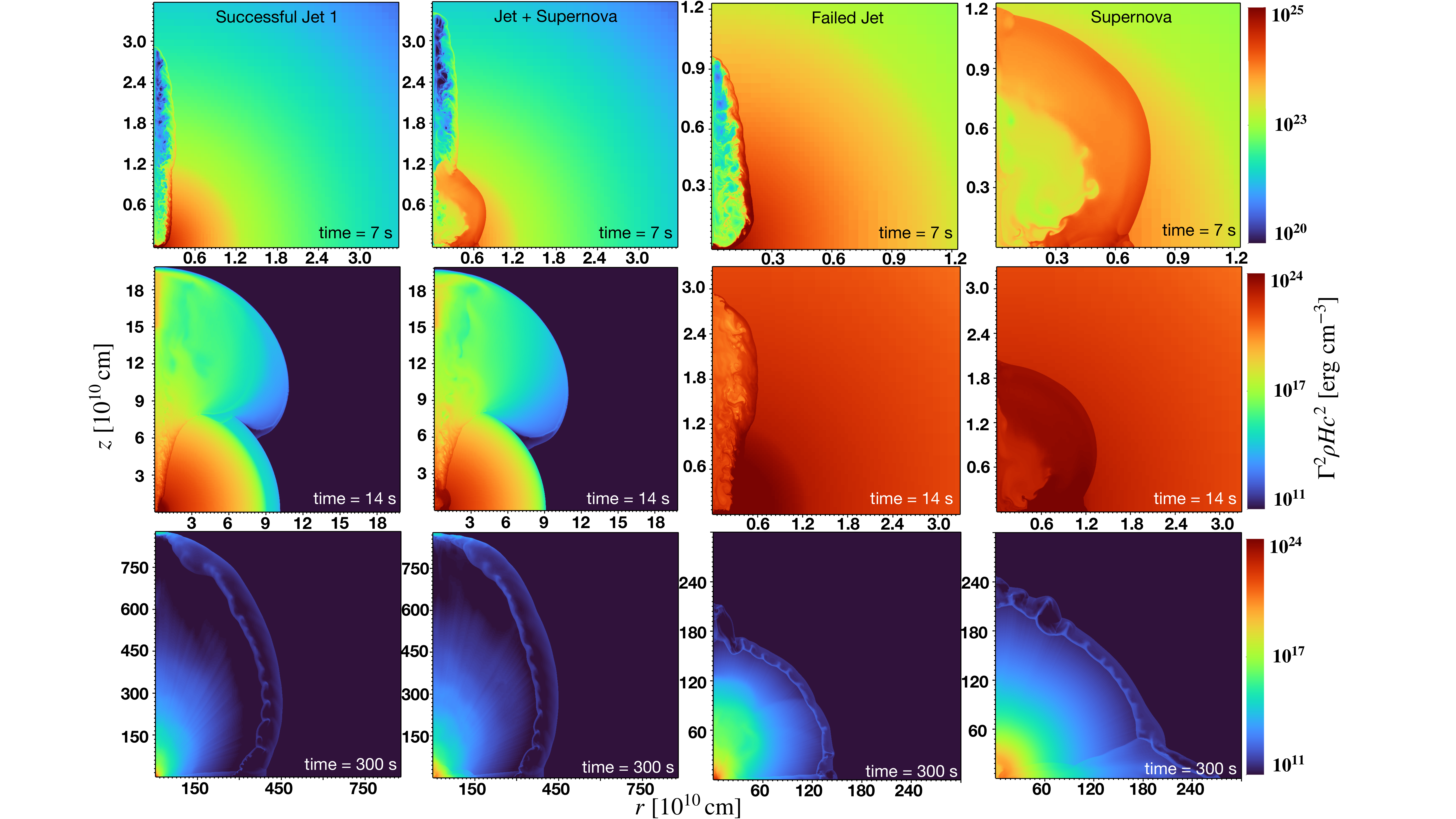}
    \caption{
    Two-dimensional plots (in cylindrical coordinates, in the r-z plane) of the energy density $\Gamma^2\rho H c^2$ . {\em Left to right panels:} successful jet, jet associated to a supernova, choked jet and SN explosion, respectively. {\em Top to bottom panels:} different evolutionary phases of the system, corresponding to 7 s (when the jet is propagating inside the progenitor star), 14 s (when successful jets have broken from the stellar surface) and 300 s (at the end of simulation).
    }
    \label{fig2}
\end{figure*}
In the case of an on-axis observer, i.e. located along the $z$-axis, $\theta_{\rm obs} = 0$, and we recover the obvious result $\Phi = \phi$. In this case, for the symmetry of the problem, we get $h_+=h_{\times}=0$.

On the other hand, in the case of a particle moving along the $z$ axis, we have $\theta=0$, which implies $\sin(2\Phi) = 0$, $\cos(2\Phi) = 1$, and $h_{\times}=0$. Also, being $\beta = \beta_z$ in this case, we get $ \cos \theta_v = \cos \theta_{\rm obs}$, and
\begin{equation}
    \frac{\beta^2\sin^2\theta_v}{1-\beta \cos \theta_v} = \frac{\beta^2(1- \cos^2 \theta_{\rm obs})}{1-\beta \cos \theta_{\rm obs}}.
\end{equation}
This function has a maximum $\left(=2(\gamma-1)/\gamma\right)$ at $\cos \theta_{\rm obs}= \beta \gamma/(\gamma+1)$. In particular, for an ultra-relativistic flow, $\gamma \gg 1$, and the maximum ($=2$) is at $\theta_{\rm obs}^2 \sim 2/\gamma$. Thus, the GW signal determined from equation \eqref{eqn:new_h+} is weakly boosted along the direction of the observer, except for observers located nearly along the jet axis (in which case $h_+=0$ as  shown above).

In practice, the calculation of the GW signals proceeds as follows. We save a large number of snapshots of our two-dimensional, axisymmetric simulations at $t=t_i$, with $i=1,..,600$ (i.e., 600 outputs, spaced by 0.5 s, during the total integration time of 300 s). The data files include the positions $R, z$ and the size $\Delta V$ of each cell, in addition to the thermal pressure, mass density and the velocity vector. Then, we remap each cell along the azimutal $\phi$ direction. We compute the values of $h_+$ and  $h_{\times}$ (to verify that it remains $\sim 0$ at all times). Then, we compute the arrival time of the GW signal generated by that particular cell, that is,
\begin{equation}
    t_{\rm obs} = t_i - (R/c) \cos\phi \sin \theta_{\rm obs} - (z/c) \cos \theta_{\rm obs} \;.
    \label{eq:tobs}
\end{equation}
We divide the time-space in the observer frame in $N_{\rm obs}$ equally-spaced time-bins. Then, we add the contribution of a certain cell to the corresponding time bin to determine $h_+$ as a function of the observer time.

\subsection{Calculation of the amplitude spectral density}\label{sec:asd_calc}

When a GW passes through an interferometer, it produces a time-series data, i.e., a succession of data points measured at certain times. The measured data $s(t)$ is a combination of the detection noise $n(t)$ and the GW signal $h(t)$ \citep{Moore2014}:
\begin{equation}
	s(t) = h(t)+n(t) ,
	\label{signal}
\end{equation}
where $h(t)= F_+ h_+ + F_\times h_\times$, being $F_+$ and $F_\times$ the antenna response patterns. For an optimal oriented source, $F_+=1$, and $h(t)\simeq h_+$.

The sensitivity of a detector to these polarizations depends upon the relative orientations of the source and detector. The challenge in the data analysis is to separate the GW signal from the noise for a given observation.

In the frequency domain $f$, the characteristic GW strain {\bf $h_c(f)$} is defined as:
\begin{equation}
    [h_c(f)]^2 = 4 f^2 |\tilde{h}(f)|^2, 
\end{equation}
where $\tilde{h}(f)$ is the Fourier transform of the strain $h(t)$, and the noise amplitude $h_n(f)$ is:
\begin{equation}
    [h_n(f)]^2 = f^2 S_n(f), 
\end{equation}
where the function $S_n(f)$ is called the power spectral density of the noise (PSD) and the signal noise ratio (SNR) can be defined by:
\begin{equation}
{\rm SNR} = \int_{0}^{\infty} df \frac{4 |\tilde{h}(f)|^2}{S_n(f)}\;.
\label{eqn:SNR_c}
\end{equation}
This characteristic strain for an astrophysical source is the amplitude of the wave times the square root of the number of periods observed. Furthermore, the amplitude spectral density (ASD) is computed as
\begin{equation}
    ASD = \sqrt{h_c(f) f^{-1/2}} = 2 f^{1/2} |\tilde{h}(f)| \;.
    \label{eq:17}
\end{equation}
The ASD is a crucial element for characterizing the detection strain during the data analysis.

The ASD and SNR are computed in this paper by considering the strain $h(t)$ computed as described in section \ref{sec:2.2}, by computing the Fourier transform and by applying equations  
\eqref{eqn:SNR_c} and \eqref{eq:17}.

The SNR for binary black holes detected by the LIGO/VIRGO network is between 6 and 26, with most events detected with a SNR of 10-20\footnote{See, e.g.,  \url{https://www.gw-openscience.org/eventapi/html/allevents/}}. In this paper we consider a conservative value SNR = 10 as detectability limit of the GW signal computed from a template-based analysis.

\section{Results}\label{sec:results}

\subsection{Jet dynamics}

 In this section, we describe the dynamics of the system for the different numerical simulations. Figure 
 \ref{fig2} shows three different evolutionary times (at 7 s, 14 s and 300 s from the top to the bottom panels) for, from left to right, a successful jet without and with an associated SN (models ``successful jet 1'' and ``jet + supernova'', for the choked jet (the ``failed jet'' model) and for a SN-like explosion (the ``supernova'' model). The ``successful jet 2'' model is qualitatively similar to the ``successful jet 1'' model (although the jet breaks out on a shorter timescale, as we will discuss below) and it is not shown in the figure.

\begin{figure}
    \centering
    \includegraphics[scale=0.6]{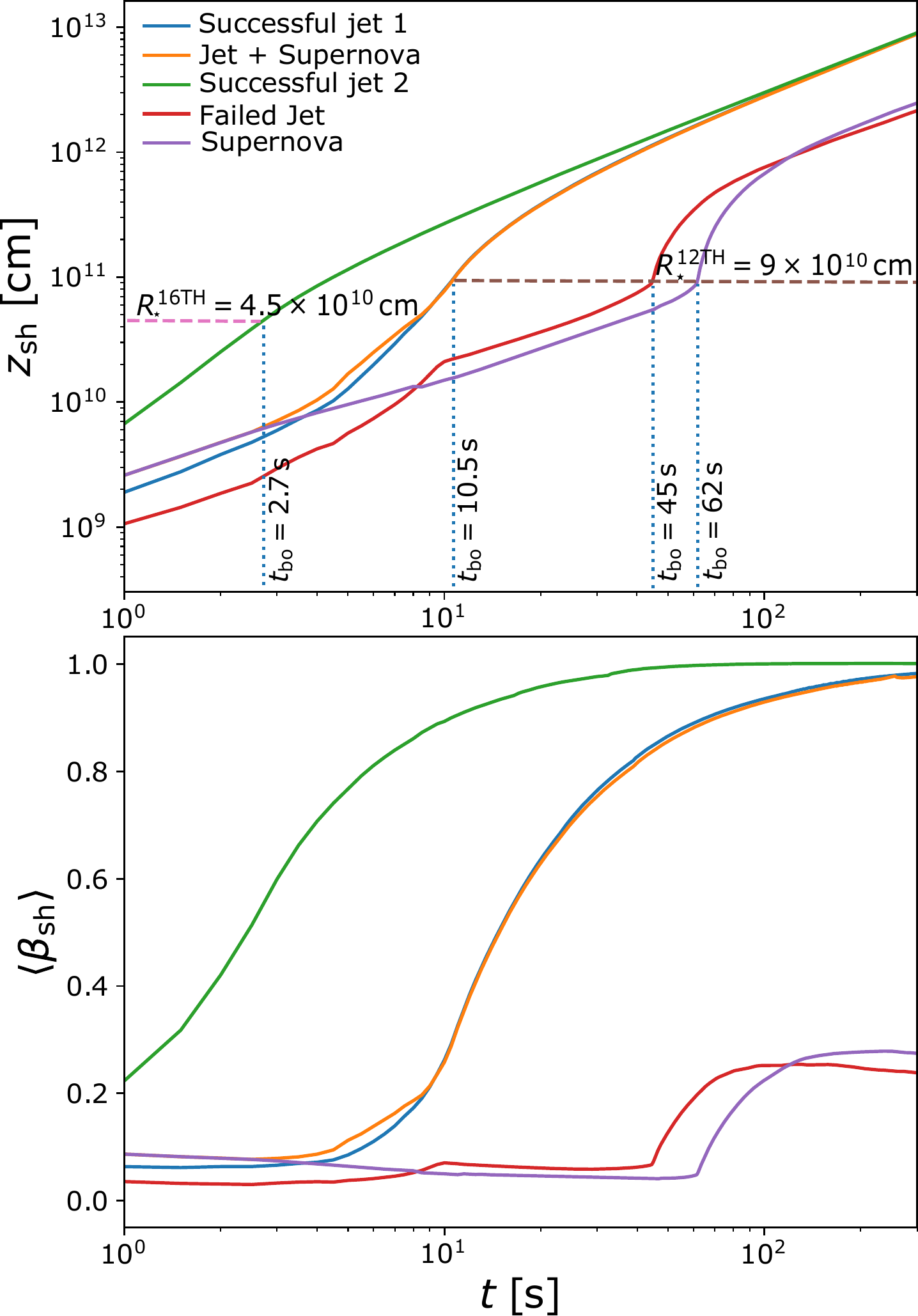}
    \caption{\emph{Top panel:} Position of the head of the jet and supernova models (as indicated by the labels) as a function of time. The horizontal dotted lines represent the radius of the star for the progenitor 16TH ($R_\star^{\rm 16TH}=4.5\times 10^{10}$ cm) for the ``successful jet 2'' model (i.e., the jet with an energy $E_{\rm jet}=10^{52}$ erg) and 12TH for all other models (with a radius $R_\star^{\rm 12TH}=9\times 10^{10}$ cm). The vertical dotted lines refer to the time in which the jet or SN head break out from the progenitor star. \emph{Bottom panel:} Average shock velocity in units of speed of light $c\;$, as a function of time.}
    \label{fig3}
\end{figure}

As shown in Figure \ref{fig2} (top panels), the ``successful Jet 1'' and ``jet + supernova'' models expands through the stellar material. At the shock front, the stellar material is heated and accelerated by the forward shock, while (in the lab frame) the jet material, launched from the central engine and propagating through the jet channel, is heated and decelerated by the reverse shock. The hot, entropy rich post-shock material expands sideways into the progenitor star, producing an extended cocoon \cite[see, e.g.,][]{Bromberg2011, Gottlieb_2018}, which helps collimating the jet. Despite this extra collimation, the jet velocity is sub-relativistic while the jet moves through the star (see Figures \ref{fig2} and \ref{fig3}).

Once the jet breaks out from the stellar surface (Figure \ref{fig2}, for the ``successful jet 1'' and ``jet + supernova'' models), the cocoon expands laterally quickly engulfing the low density region surrounding the progenitor star, while the entropy rich material, close to the jet axis, accelerates converting thermal to kinetic energy. The cocoon material remains strongly stratified both along the radial and the polar direction, moving at mildly relativistic speeds (close to the jet axis) and sub-relativistic speeds close to the equatorial plane.

Once the jet expands to larger distances (Figure 
\ref{fig2}, left-bottom panel), the fast moving material remains confined into a thin shell with size $\gtrsim t_j c$ ($\sim 3\times 10^{11}$ in the successful jet simulations shown in the figure), where $t_j$ is the time during which the jet is injected by the central engine. On the other hand, the cocoon begins to decelerate, specially close to the equatorial plane where the cocoon energy is lower, as indicated by the presence of Rayleigh-Taylor instabilities visible in Figure \ref{fig2}.

The simulation of the jet associated to a SN (the ``jet + supernova'' model) is qualitatively similar to the one without the SN (the ``successful jet 1'' model). In this simulation, the jet is launched with a delay of 1 s with respect to the SN. After a few seconds, the jet head reaches the SN shock front, breaking out of it and expanding through the progenitor star. The late phases are also similar to the case of a jet without a SN discussed above, except that, at large times, the SN shock front breaks out from the progenitor star into the jet cocoon.

We notice that the general outcome of the system depends on the time when the jet breaks out from the SN. If, for instance, the jet energy, opening angle and duration are such that the SN shock front breaks out first from the stellar surface, then the jet will remain trapped  inside the expanding SN, depositing its energy in the deep layers of the SN ejecta. The result of the interaction between the SN, the jet and its cocoon leads to a rich landscape of scenarios which have not been studied in detail yet \citep[see][for a qualitative description]{DeColle2021}.

The third column of Figure \ref{fig2} shows the case of a choked jet (the ``failed jet'' model). In this case, the jet opening angle is larger by a factor of $\sim$ 2, so that the luminosity per unit solid angle drops by a factor of $\sim 4$. Then, the jet duration (10 s) is not large enough for the jet to  break through the progenitor star. Once the jet power is switched off, the relativistic moving material crosses the jet channel in a time $R_h/c\sim \beta_h t_j$, being $R_h$ and $\beta_h \sim 0.1-0.3$ c the head position and velocity, and $t_j$ the jet injection time. Once all the jet material arrives to the head of the jet, the jet quickly expands laterally and decelerate. Then, it can break out from the stellar surface into a more spherical explosion (see the bottom panel of the figure).

The last column of Figure \ref{fig2} shows a nearly spherical explosion, qualitatively representing a SN explosion (the ``supernova'' model). In this case, the shock breakout is also nearly spherical. Nevertheless, we notice that realistic 3D simulations of SN explosions show a much more asymmetric, turbulent behaviour not captured in these 2D simulations.

Figure \ref{fig3} shows the evolution of the head of the jet ($z_{\rm sh}$ hereafter) and its average velocity, as a function of time, for the different models. As discussed above, the velocity of the shock front is sub-relativistic inside the progenitor star. Once the shock front approaches the stellar surface, it quickly accelerates due to the large density gradients. This is visible both in the top panel of Figure \ref{fig3}, where the slope of the curves showing $z_{\rm sh}$ vs $t$ becomes steeper just after the breakout (represented by the vertical dotted lines), and in the bottom panel, where the average velocity increases quickly after the breakout. Then, the SN and the choked jet cases achieve a velocity of $\sim 0.2$ c, while the successful jets (with or without SN associated) continue accelerating until the end of the simulation. As mentioned before, the acceleration process is related to the conversion of thermal to kinetic energy. At the end of the process, the jet head will arrive to a terminal Lorentz factor $\Gamma_j\sim E_j/M_j c^2 \gg 1$.

Finally, we notice that the high luminosity model (``successful jet 2'') is qualitatively similar to the ``successful jet 1'' model, with the main difference being the timescales for the different phases to occur. As the luminosity is larger, the jet duration is shorter, and the progenitor star is smaller, the jet will break out from the stellar surface in a much smaller time, and it will accelerate faster to its final velocity (see Figure \ref{fig4}).

\subsection{GW emission}

\begin{figure}
    \centering
    \includegraphics[scale=0.6]{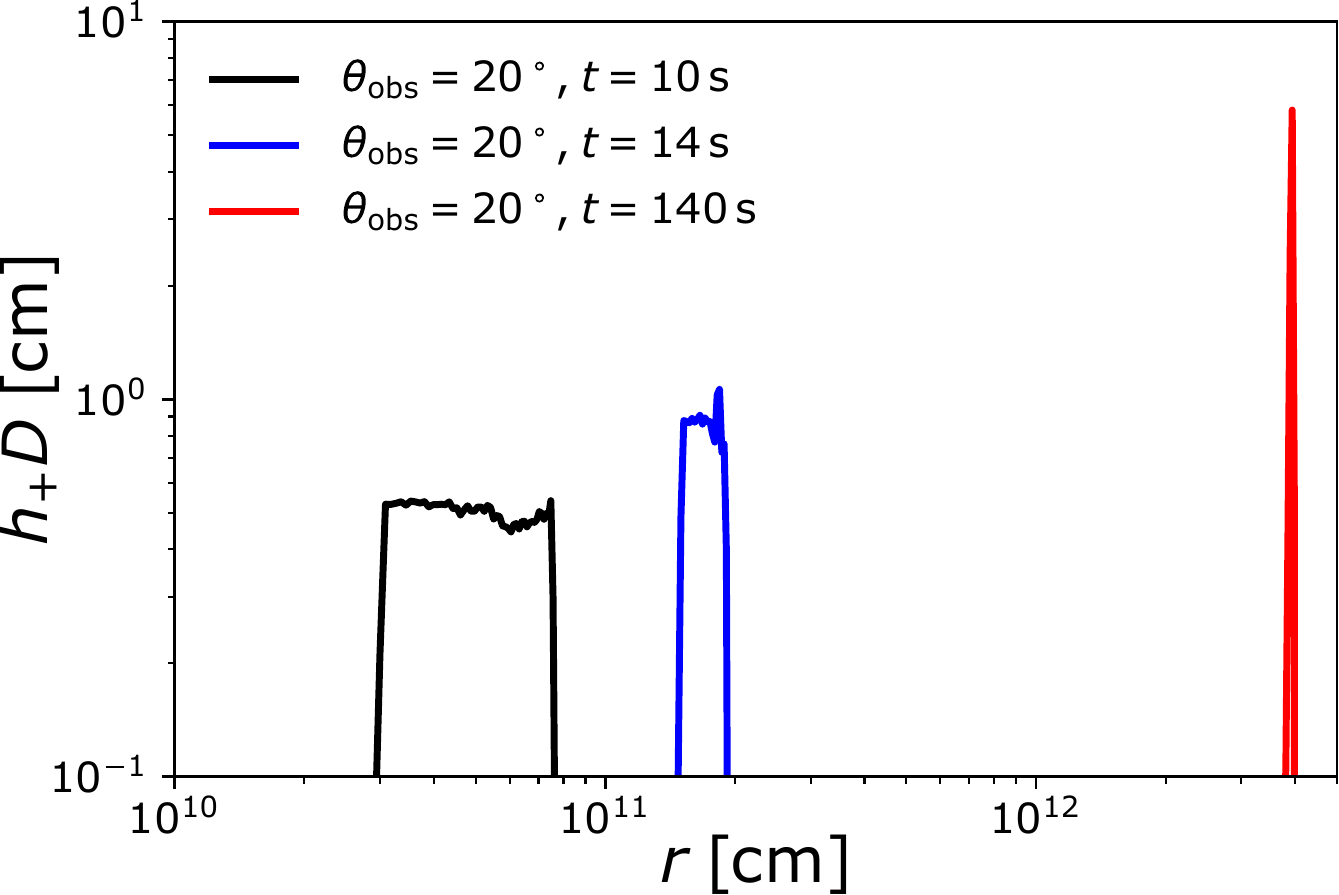}
    \caption{GW signal $h_+$ as a function of $z$ extracted from the ``jet+supernova'' model, corresponding to $t=10$ s, $t=14$ s and $t=20$ s (in the lab frame). The observer is located at $\theta_{\rm obs} = 20^\circ$. The figure shows that the GW signal is generated along all the jet channel (black, blue curves) at early times, and in a thin shell at large times (red curve), corresponding to the location of the highly relativistic material.}
    \label{fig4}
\end{figure}

To understand where the GW signal originates from, we show in Figure \ref{fig4} the amplitude of the GW signal $h_+$ as a function of $z$, at different times, i.e., integrating over the radial and azimuthal directions. During the first 10 s, the jet is continuously injected into the computational box, and the jet energy increases along the jet channel (see Figure \ref{fig3}). As shown by the black curve, corresponding to $t=10$ s, the GW signal  is produced along most of the jet channel. The small fluctuations correspond to the presence of recollimation shocks. As the jet pressure is larger than the cocoon pressure, the jet expands laterally into the cocoon, until when both pressures are approximately equal. Then, a recollimation shock is created, pinching the jet onto the jet axis. This produces strong fluctuations in the jet velocity and energies, which lead to the observed fluctuations in the GW signal seen in Figure \ref{fig4}.

Once the jet breaks out from the star, the energy and velocity into the emitting region becomes more uniform. As discussed above, the jet velocity increases strongly achieving a  Lorentz factor close to the terminal value (set to 100 in the simulation, see section \ref{sec:methos}). While a fraction of the total energy is stored in the cocoon, the cocoon does not contribute significantly to the GW signal, as it moves at most at mildly relativistic speeds. This can be seen in the red curve shown in Figure \ref{fig4} (corresponding to $t=140$ s), in which it is evident that the region emitting the GW signal is limited to the fast moving jet material.

\begin{figure}
    \centering
    \includegraphics[scale=0.6]{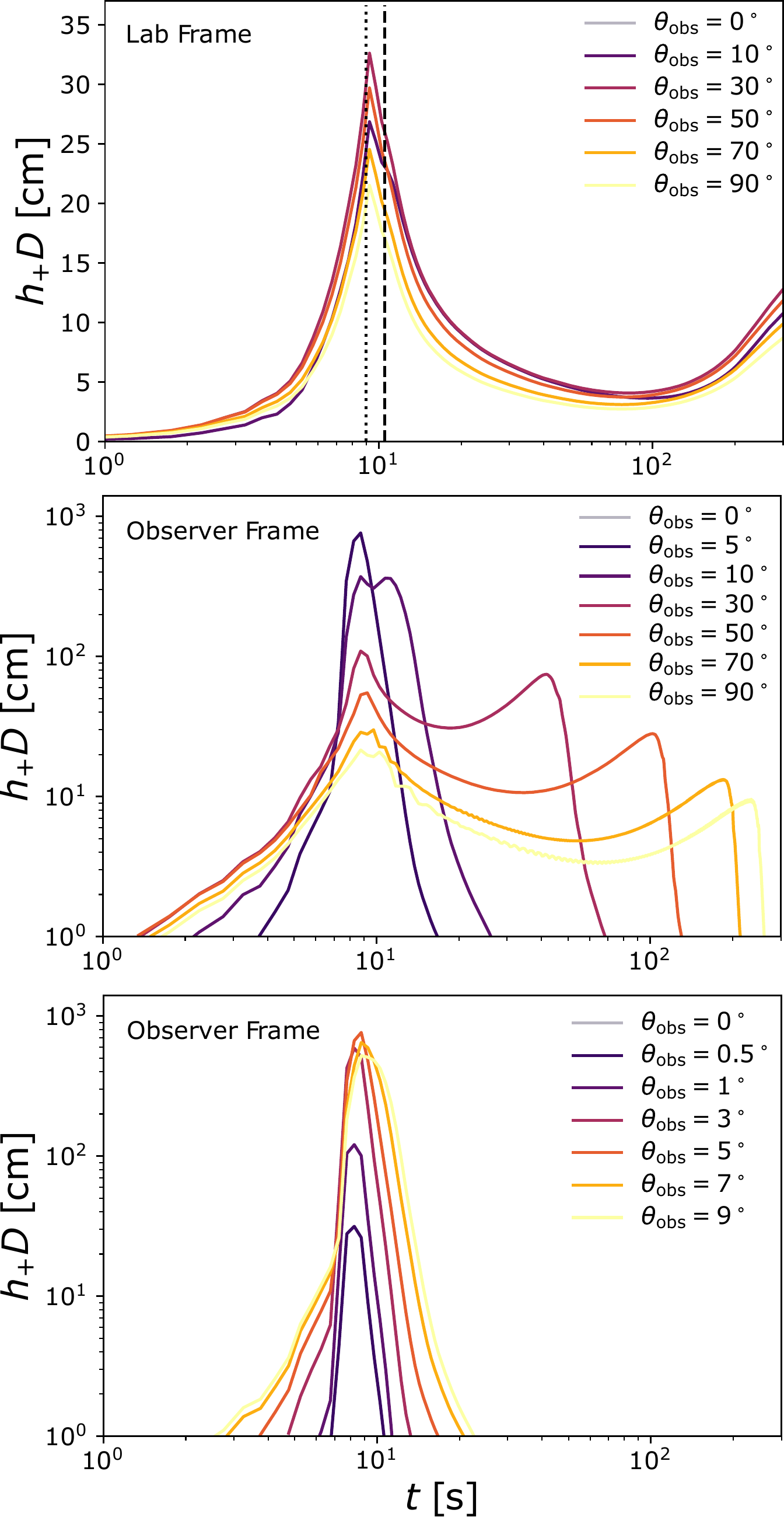}
     \caption{GW strain (multiplied by distance $D$) as a function of the lab frame time (top panel) and the observer time (center, bottom panels). The different curves correspond to different observer angles, ranging from 0$^\circ$ to 90$^\circ$ in the top panel, center panels, and from 0$^\circ$ to 9$^\circ$ in the bottom panel. The calculations correspond to the case of a successful jet with a duration of 10 s without any associated SN (model ``successful jet 1). The vertical dotted and dashed lines in the top panel refer to the jet injection time (9 s) and the jet break out time ($\sim$10.5 s).}
    \label{fig5}
\end{figure}

Figure \ref{fig5} shows $h_{+} D$ as a function of time. $h_{\times} D$, not shown in the figure, remains close to zero (at machine precision) at all time, given that all simulations are axisymmetric. To illustrate the effect of the arrival times on the shape of the GW signal, we show the GW amplitude in the lab frame (top panel), i.e., computed assuming $t_{\rm obs} = t$ in equation \eqref{eq:tobs}, and in the observer frame (center, bottom panels) for the successful jet model without an associated SN.
In the lab frame, the GW signal presents two peaks, the first one at $t=t_j$, i.e., corresponding to the time when the jet power is switched off from the central engine, and the second at the very end of the simulation, corresponding to the acceleration of the jet to its terminal velocity.

Equation \eqref{eqn:new_h+} implies that a constantly powered jet with constant velocity (along the $z$-axis) and $E_j=L_j t$, with also $L_j(t)=L_j$ constant, would produce a GW signal increasing linearly with time (see also \citealt{Yu_2020}). 
Figure \ref{fig5} shows that the increase before the first peak is not linear, due to the jet acceleration as it approaches the stellar surface and it moves through a thinner medium (see Figure \ref{fig3}, bottom panel). As soon as the jet luminosity starts dropping\footnote{The jet injection time is $t_j=10$ s, but, to avoid numerical problems related with the strong rarefaction wave produced once the jet is switched off, we set a jet luminosity dropping linearly between 9 s and 10 s.} at $t=9$ s, the GW amplitude quickly drops with time. At larger distances from the central engine, the GW amplitude increases again due to the acceleration of the jet material. Once the jet achieves its terminal velocity, that is, after transforming most of its thermal to kinetic energy, the GW amplitude achieves a second peak before dropping again with time. Unfortunately, the second peak is not completely resolved in our simulations, as it happens (in the lab frame) at times larger than the simulated 300 s. Then, the value of the GW signal at the second peak should then be taken as a lower limit to the real value.
In the lab frame, the dependence on the observing angle is weak. Except for observer located exactly on the jet axis, for which $h_+=0$, there is a difference $\lesssim 2$ between the values of $h_+$ computed at different observer angles.

The central and right panels of Figure \ref{fig5} show the same calculations, but in the observer frame. A qualitative understanding of the behaviour of $h_+$ in this case can be attained by assuming that all GW signal is coming from a region very close to the jet axis. In this case, $R=0$, and equation \eqref{eq:tobs} reduces to
\begin{equation}
    t_{\rm obs} = t_n - (z/c) \cos \theta_{\rm obs} \;.
\end{equation}
Then, assuming that the emission comes from a single point source moving with constant velocity $\beta$, we get
\begin{equation}
  t_{\rm obs} = t \left(1 - \beta \cos \theta_{\rm obs}\right) \;.
\end{equation}
For observers located at large observing angles, $\theta_{\rm obs}\gg 0$, $t_{\rm obs} \sim t$ and the GW arrival time is the same as the time when the signal is produced (except of course for the time $D/c$ needed for the signal to propagate from the source to the Earth). On the other hand, for observers located at small observing angles, 
\begin{equation}
     \cos \theta_{\rm obs}\sim 1-\frac{\theta_{\rm obs}^2}{2},
\end{equation}
and 
\begin{equation}
    t_{\rm obs}\sim t \left(1 - \beta  + \frac{\beta \theta_{\rm obs}^2}{2}\right) \sim t\;\frac{1 + \Gamma^2 \theta_{\rm obs}^2}{2\Gamma^2} .
\end{equation}
Then, for 
\begin{equation}
    \theta_{\rm obs} \ll \frac{1}{\Gamma}\sim 6^\circ \left(\frac{\Gamma}{10}\right)^{-1} , 
\end{equation}
we have
\begin{equation}
    t_{\rm obs} \sim \frac{t}{2\Gamma^2},
\end{equation}
and the GW signal arrival time is reduced by a factor of a few hundred with respect to the GW signal as seen in the lab frame, while for $\theta_{\rm obs} \gg 1/\Gamma $, we have 
\begin{equation}
    t_{\rm obs} \sim \frac{t \, \theta_{\rm obs}^2}{2}.
\end{equation}

As shown in Figure \ref{fig5}, the GW signal is very different in the observer frame with respect to the lab frame. Consistently with the discussion above, the second peak moves to increasingly smaller observer times for smaller observer angles. So, at $\theta_{\rm obs} = 5^\circ$, the second peak drops substantially, overlapping the first peak. As the simulations output files are saved every 0.5 s, this implies that, for this observer angle, the two peaks are separated by less than 0.5 s., while, e.g., the second peak moves at $\sim 12$ s, $\sim 22$ s for observers located at $\theta_{\rm obs} = 10^\circ, 20^\circ$ respectively. As more GW radiation arrives during a shorter time, the amplitude of the two peaks increase substantially, specially for small observer angles.
The bottom panel shows that the maximum in the GW signal is obtained between $\theta_{\rm obs} = 3^\circ$ and $\theta_{\rm obs} = 7^\circ$, i.e., for observers located at the edge of the jet. Although it is barely visible due to the size of the bins in time (0.5 s as mentioned before), the break-out from the progenitor star produces a small change in the slope of the curves.

\begin{figure}
    \centering
    \includegraphics[scale=0.6]{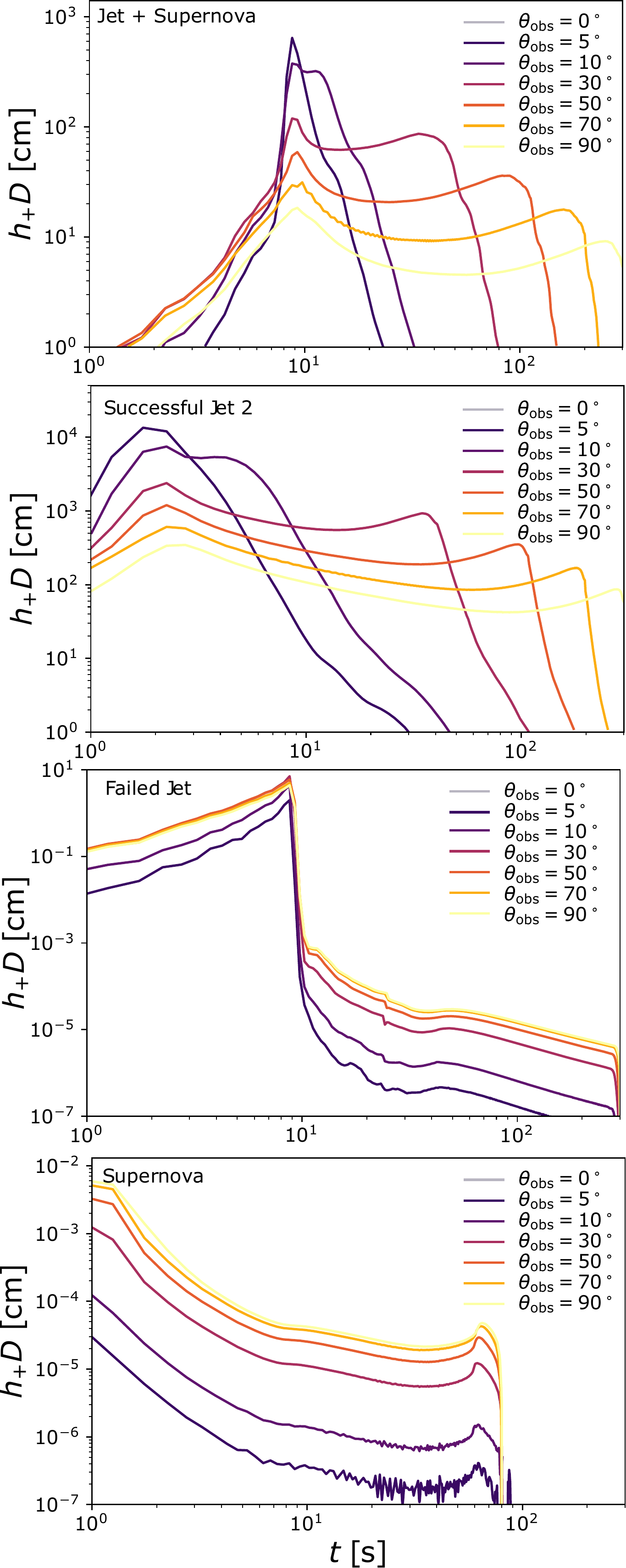}
    \caption{GW strain as a function of the observer time for the models considered in the paper. \emph{From top to bottom}: successful jet associated to a SN,  successful jet with a shorter duration and moving through a more compact star, choked jet and SN model. 
    The different models are computed at different observer angles $\theta_{\rm obs}$.}
    \label{fig6}
\end{figure}

Figure \ref{fig6} shows the GW amplitude $h_+ D$ for the other models considered. The ``successful jet 1'' and ``jet + supernova'' models produce similar results (compare the upper panel of Figure \ref{fig6} with the middle panel of Figure \ref{fig5}). The GWs produced by the luminous, ``successful jet 2'' shown in the second panel also presents a similar behaviour, but with peaks located at shorter times, and a much larger amplitude at peak ($\sim 13000$ cm vs $\sim 650$ cm). In the case of the ``failed jet'', $h_+$ increases for $t\leq t_j$, to then drop on a short timescale ($\lesssim 0.5$ s). The peak achieved for this model is $\sim 2-3$ order of magnitude smaller than in the other cases.
Finally, the GW signal produced by a {\rm SN} is several orders of magnitude smaller, as the velocity of the SN shock front remains always sub-relativistic. Anyway, we note that our simulations do not capture the initial, larger GW signal produced by the early propagation of the SN shock front immediately after the collapse, because we follow the propagation far away from the central engine.
\begin{figure*}
    \centering
    \includegraphics[scale=0.7]{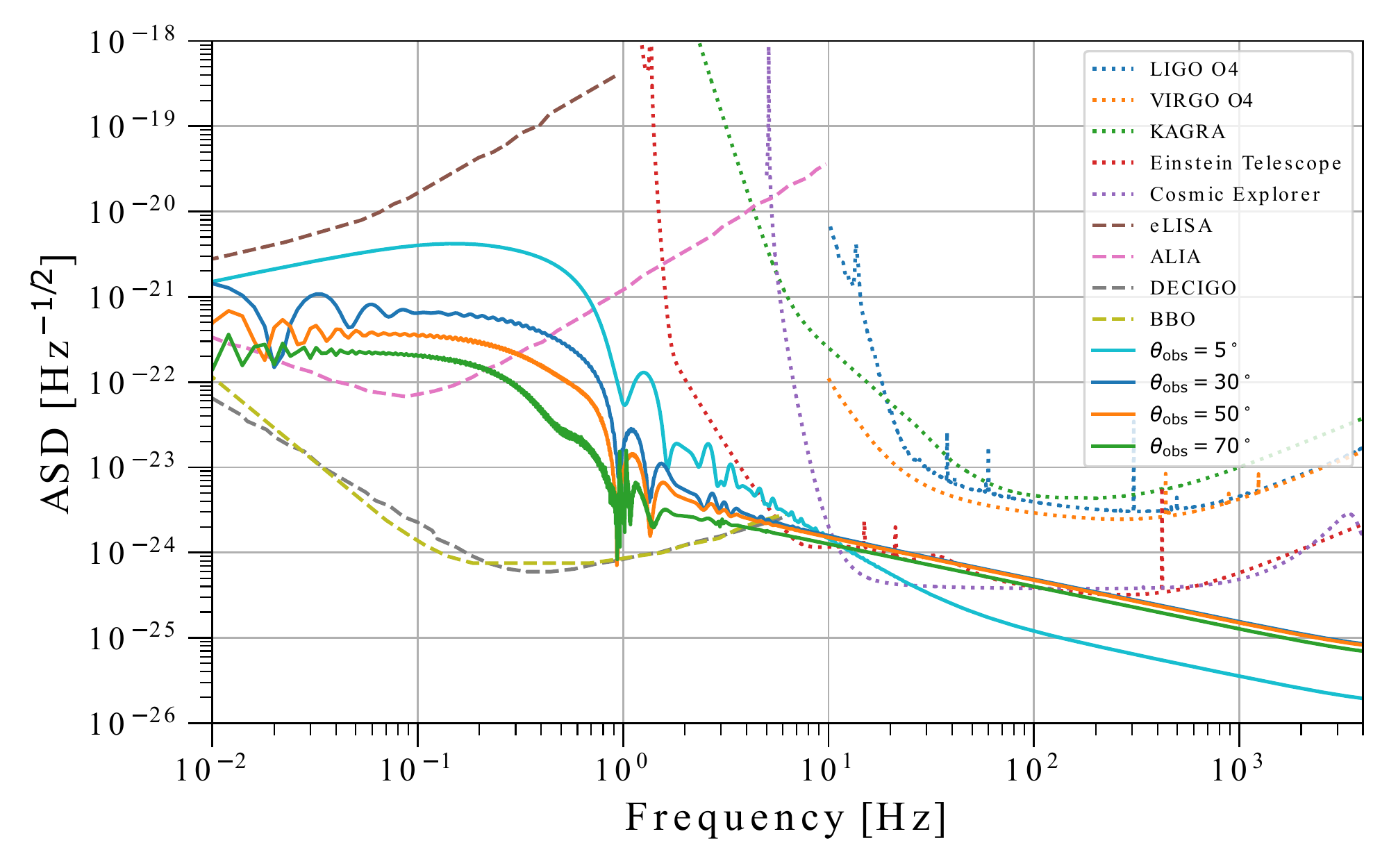}
    \caption{Amplitude spectral density (ASD) of GW signal computed from the ``successful jet 2'' (lasting $t_j=2.5$ s) at $D=1\,$Mpc, and the ASD of the noise floor for
    LIGO 04, VIRGO 04, Kagra, the Einstein Telescope, eLISA, DECIGO, the Big-bang Observatory (BBO) and the Advanced Laser Interferometer Antenna ALIA. Dotted lines refer to ground-based interferometers, while dashed lines refer to space-based interferometers. The detection limits were taken from \citet{Moore2014}.
    }
    \label{fig7}
\end{figure*}

\begin{table*}
    \centering
    \begin{tabular}{cccccccc}
        \hline
               Detector  & \multicolumn{2}{c}{SNR  } & \multicolumn{2}{c}{Distance [Mpc]} & \multicolumn{3}{c}{Rate [yr$^{-1}$]} \\ 
                 & $5^\circ$ & $70^\circ$ & $5^\circ$ & $70^\circ$& $0^\circ-10^\circ$ & $10^\circ-40^\circ$ & $40^\circ-90^\circ$  \\ \hline 
                 
        LIGO O4  & $3.8\times 10^{-3}$ & $1.3 \times 10^{-2}$ & $1.5\times 10^{-2}$ & $5.1\times 10^{-2}$ & $1.5 \times 10^{-12}$ & $1.9\times 10^{-10}$ & $4.2\times 10^{-10}$ \\ \hline

        VIRGO O4    &  $2.0\times 10^{-3}$ & $5.5 \times 10^{-3}$ &  $2.2\times 10^{-2}$  & $2.2 \times 10^{-2}$ & $7.3 \times 10^{-13}$ & $1.8\times 10^{-11}$ & $3.6\times 10^{-11}$ \\ \hline
        
        KAGRA    &  $8.9\times 10^{-3}$ & $2.8\times 10^{-3}$ &  $7.3\times 10^{-3}$  & $2.3\times 10^{-2}$ & $1.6\times 10^{-14}$ & $2.1\times 10^{-12}$  & $5.0\times 10^{-12}$ \\ \hline
        
        Einstein Telescope &   $4.4\times 10^{-2}$ & $6.2\times 10^{-2}$ & $3.5\times 10^{-1}$ & $5.0\times 10^{-1}$ & $3.9\times 10^{-10}$ & $2.3\times 10^{-8}$ & $5.3\times 10^{-8}$ \\ \hline
        
        Cosmic Explorer    & $3.8\times 10^{-2}$ & $6.7\times 10^{-2}$ & $3.0\times 10^{-1}$ & $5.3\times 10^{-1}$ & $3.4\times 10^{-10}$  & $2.8 \times 10^{-8}$ & $6.4 \times 10^{-8}$ \\ \hline
         
        eLISA  & $2.1\times 10^{-2}$ & $3.9\times 10^{-3}$ & $8.5\times 10^{-2}$ & $1.5\times 10^{-2}$ & $5.5 \times 10^{-11}$ & $3.7\times 10^{-10}$ & $4.0\times 10^{-11}$  \\ \hline 
 
        ALIA   & $1.6$ & $9.3\times 10^{-2}$ & $6.4$ & $3.7\times 10^{-1}$ & $1.3\times 10^{-5}$ & $1.2\times 10^{-5}$ & $4.5\times 10^{-7}$ \\ \hline
         
        DECIGO & $1.5\times 10^{2}$ & $4.7$ & $6.0\times 10^2 $ & $1.8\times 10^{1}$ & $7.5$ & $2.2$ & $1.0 \times 10^{-1}$ \\ \hline
       
        BBO    & $1.5\times 10^2$ & $5.4$ & $6.0\times 10^2$ & $2.1\times 10^1$ & $ 7.9$ & $2.5$ & $1.2\times 10^{-1}$ \\ \hline

    \end{tabular}
    \caption{The columns refer to:
the observatories considered (see Figure \ref{fig7}), the signal-to-noise ratio (SNR) for a jet seen at an observer angle $\theta_{\rm obs}=5^\circ, 70^\circ$ and at a distance of 40 Mpc, the distance where SNR = 10, and the number of events detected per year along different solid angles. The values refer to the ``successful jet 2'' model. }
    \label{tab:SNR}
\end{table*}

\section{Discussion}\label{sec:discussion}

In this paper, we have presented numerical simulations of the propagation of relativistic jets through a massive, progenitor star, the break-out and the expansion of the jet up to distances $\sim 10^{13}$~cm, and computed the resulting GW signal as a function of the observer angle.

Previous studies of GW memory from GRB jets have focused on the neutrinos produced by the central engine during the jet formation \citep{Hiramatsu2005,Suwa2009,Kotake2012}, on internal shocks and shock deceleration during late stages of evolution \citep{Akiba2013} and on the jet acceleration \citep{piran13,Yu_2020,piran21}. These studies have used an analytic description of the jet, often taken as an accelerating point mass. In our study we compute the GW signal by using the dynamics of the jet while it crosses the progenitor star and it accelerates through the circumstellar medium.
Although our results qualitatively confirm previous findings, our numerical simulations allow us to give a quantitative prediction of the expected GW signal. 

\citet{Akiba2013} showed that the GW signal computed during the shock deceleration is about $\sim 1000$ times smaller than the one determined by our simulations, although we sample different distances, with our simulations extending up to $10^{13}$ cm, while \citet{Akiba2013} studied the propagation of the jet during the prompt emission, i.e. at $R_{\rm sh} \sim 10^{13}-10^{15}$ cm.

\citet{piran13,piran21} studied the acceleration of the jet up to ultra-relativistic speeds. They showed that the jet acceleration produces a peak in the GW signal, which depends on the observer angle.
Their study can be applied, in our context, to the acceleration of the jet when it breaks out from the star. Thus, the peak they observe in their calculations is equivalent to the second peak seen in Figure \ref{fig5} and \ref{fig6}.

\citet{Yu_2020} employed an analytical model for the dynamics of the jet through the progenitor star (applying it also to sGRBs). They computed the acceleration of the shock front as it approaches the stellar surface. Although the results are qualitatively similar, the temporal evolution of $h_+D$ is different (compare, e.g., their Figure 3 with our Figures \ref{fig5} and \ref{fig6}). As they mention, observing the GW signal would probe the jet propagation and the interior of the progenitor star. Nevertheless, we argue in this paper that numerical models are needed to get a proper quantitative prediction.

The GW signal is ``anti-beamed'' \citep{Segalis_2001,Sago_2004,piran13,piran21}. Nevertheless, we notice that the GW signal is strongly suppressed only for observer located at $\theta_{\rm obs} \approx 0^\circ$. As shown in the bottom panel of Figure \ref{fig5}, it increases for larger observer angles (respect to the jet opening angle $\theta_j$), peaking at $\theta_{\rm obs}\sim \theta_j$ (e.g., the GW signal is $\sim$ 1/2 of the peak at $\theta_{\rm obs} = \theta_j/2$). 
In contrast with the prediction obtained by considering analytical models, then, we expect to see GWs associated to GRBs seen nearly on-axis. Also, we expect than in three-dimensional numerical simulations, in which the symmetry with respect to the main axis of propagation of the jet is broken, the propagating jet would produce a GW signal also on-axis.

The other clear feature resulting from our models is the presence of a double peak structure in the GW signal, due to two characteristic acceleration phases: a) inside the progenitor star, as the jet move through a lower density medium as it approaches the stellar surface; and b) after the breakout, as the jet accelerates converting thermal to kinetic energy. The timescales of the two peaks reflect directly the duration of the jet $t_j$ (the first peak) and the observer angle (with larger timescales corresponding to larger $\theta_{\rm obs}$, see Figures \ref{fig5} and \ref{fig6}). 

As discussed above, the slope of the GW signal before and after the first peak (see, e.g., Figure \ref{fig6}) depends on the stellar structure and on the jet luminosity. For instance, we can expect a shallower increase for a jet with a luminosity decreasing with time. Thus, GW observations by future detectors may provide direct information on the central engine activity (e.g., jet duration and luminosity history), the stellar structure, the observer angle and the acceleration process after breakout.

Figure \ref{fig7} shows the amplitude spectral density computed from the numerical simulation of the ``successful jet 2'' model, by employing the methods described in Section \ref{sec:asd_calc}. In the figure, we can observe the range of frequency $10^{-2}-10^{3}$ Hz and the ASD $10^{-26}-10^{-10}$ Hz$^{-1/2}$ for several interferometers, and for the astrophysical signal analyzed in our study.

LIGO-VIRGO detectors were the first-generation detectors. They have completed science runs O1, O2, O3. They are currently being upgraded for O4 which will start to take data during February 2023. The KAGRA \citep{Aso2013eba} interferometer detector will join the LIGO/VIRGO collaboration during 2023. Future interferometer include \citep{Moore2014} the Laser Interferometer Space Antena (eLISA), the Advanced Laser Interferometer Antenna (ALIA) \citep{Sathyaprakash:2009xs}, DECIGO, the Big Bang Observer (BBO, \citealt{Yagi:2011wg}), and the  Einstein Telescope (ET)/Cosmic Explorer (CE) \citep{Hild:2010id}. The ASD for all these interferometers are included in Figure \ref{fig7}.

Figure \ref{fig7} shows the ASDs computed from the simulation assuming a GRB jet at 1 Mpc. The signal peaks at low frequencies ($\sim 0.1$ Hz), and depends strongly on the observer angle, with a peak between $5\times 10^{-21} (D/1\; {\rm Mpc})^{-1}$ at $\theta_{\rm obs} = 5^\circ$ and $2\times 10^{-22} (D/1\; {\rm Mpc})^{-1}$ at $\theta_{\rm obs} = 70^\circ$. At larger frequencies, the signal drops to much smaller values, being $\sim$ one order of magnitude below the ASD of LIGO/VIRGO. However, our times series is sampled each 0.5 s, corresponding to a maximum frequency of 2 Hz, so that results above this frequency should be taken carefully.

In table \ref{tab:SNR} we estimate the detectability of the ``successful jet 2'' model (i.e., a relativistic jet with a total energy of $10^{52}$ erg lasting 2.5 s), considering a distance of 40 Mpc (the second and third columns of table \ref{tab:SNR}) using equation \ref{eqn:SNR_c}, for present and planned interferometers (first column) , at two characteristic observer angles ($\theta_{\rm obs}=5^\circ, 70^\circ$). The SNR is very low for ground-based interferometers ($\lesssim 4.4\times 10^{-2}$), is $\approx 1$ for ALIA and $\gg 1$ for DECIGO and BBO for a nearly on-axis observer (at $\theta_{\rm obs}=5^\circ$), and drops to smaller values for off-axis observers.

The third and fourth columns of table \ref{tab:SNR} show the distance (in Mpc) where SNR = 10, by using the relation Distance = (SNR$_{40 \; \rm Mpc}$/10) $\times$ 40 Mpc\footnote{It is easy to rescale the detectability range for different SNR thresholds as the SNR is inversely proportional to distance.}. Only galactic GRBs can be detected (while crossing the progenitor star) by LIGO/VIRGO (with a SNR=10 at $1.5-5.1\times10^{-2}$ Mpc = 15-51 kpc depending on $\theta_{\rm obs}$) and Kagra (with a SNR=10 at $7.3-23\times10^{-3}$ Mpc = 7.3-23 kpc), while DECIGO and BBO can detect GRBs with an SNR=10 up to 18-600 Mpc depenging on the observer angle.

The (uncertain) expected GRB rate is 100-1000 Gpc$^{-3}$ yr$^{-1}$ \citep[see, e.g.,][]{Fryer2002,WandermanPiran2010,Cao2011,Abbott17c}.
The sixth and seventh columns of table \ref{tab:SNR} show the expected GRB/GW detection rate by assuming an (optimistic) GRB rate of 1000 Gpc$^{-3}$ yr$^{-1}$. We compute the volume corresponding to a SNR of 10 for each solid angle, and the expected GRB rate within this solid angle\footnote{This is an order magnitude estimation. A more precise calculation would require to include the GRB energy and time duration distribution. We leave it for a future study.}. 
The expected rate is very low for ground-based interferometers, while $\sim 8$ LGRB jets per year are expected to be detected by future spaced-based interferometers at small observer angles ($\theta_{\rm obs} \lesssim 10^\circ$), and $\sim 2$ LGRB jets per decade for GRB jets observed at $\theta_{\rm obs} =  40-90^\circ$.

In agreement with previous estimates \citep{Sago_2004,Hiramatsu2005,Suwa2009,Kotake2012,Sun2012,Akiba2013,piran13,Du_2018,Yu_2020,piran21}, 
the LGRB memory from jets crossing the progenitor stars are expected to be undetectable with LIGO/VIRGO and KAGRA. 
Given the (uncertain) expected GRB rate of 100-1000 Gpc$^{-3}$ yr$^{-1}$ \citep[see, e.g.,][]{Fryer2002,WandermanPiran2010,Cao2011,Abbott17c}, the GW memory from jet/shock
propagation in very rare galactic GRB jets is eventually detectable with LIGO/VIRGO.
Future space-based low-frequency instruments, as DECIGO and BBO, will easily detect the GW memory from GRB jets located up to distances $\lesssim 600$ Mpc, as shown Table \ref{tab:SNR}.

In addition to successful jets, producing the observed gamma-ray emission, other high energy transients are likely associated to a central engine activity and to the propagation of a relativistic jets, including  low-luminosity GRBs \citep{Campana2006,Soderberg2006,Starling2011,Margutti2013}, relativistic SNe \citep{Soderberg2010,Margutti2014,Milisavljevic2015}, and X-ray flashes \citep{pian2006,BrombergNakarPiran2011,NakarSari2012}. In addition, it has been suggested that SNe (in particular, broad-line type Ic) could be produced by the propagation of a choked jet \citep[e.g.,][]{Piran2019,soker2022}.

These events could be detectable at shorter distances. Our results show that the GW strain depends mainly on the jet luminosity and the jet velocity.
Jets choked while deep inside the progenitor stars, as the one simulated in this paper, will have a very low signal (see Figure \ref{fig6}, third panel) as their velocity is only mildly relativistic when the jet is switched-off from the central engine. Nevertheless, jets lasting for longer times, i.e. arriving closer to the stellar surface before being choked, will accelerate to relativistic speeds producing signals similar to those of successful jets. The quoted detection distances may also be optimistic, if template-based searches cannot be used (and, consequently, the SNR threshold for detection
is raised).

Finally, we notice that, while we have simulated relativistic jets leading to LGRBs (i.e., associated to the collapse of massive stars), a similar outcome is expected for SGRBs, associated to the coalescence of massive stars. These jets are expected to last for shorter times, to have smaller total energies and can move through smaller density media, so than they could achieves relativistic velocities on shorter timescales. Detailed numerical simulations are needed to understand whereas the expected signal would be larger for jets associated to LGRBs or SGRBs.

\section{Conclusions}\label{sec:conclusions}

In this paper, we have presented numerical simulations of relativistic jets associated to long GRB. We have computed the resulting GW signal for successful jets, choked jets, and jets associated to a SN. In successful jets (accompanied or not by a SN), the GW signal is characterised by a double peak structure, with amplitudes $h_+ D$ ranging from hundreds to several thousand. The first peak corresponds to the jet injection from the central engine, while the second peak corresponds to the jet acceleration while it breaks out from the star. In addition, the slope of the GW signals track directly the luminosity history of the GRB jets, and the structure of the progenitor star. 

As GRBs are the product of collimated jets seen nearly on-axis, given the detected GRB rate, the volumetric rate depends on the jet angle and on the jet structure. Thus, the GRB volumetric rate is highly uncertain ($\sim$ 100-1000 Gpc$^{-3}$ yr$^{-1}$). As illustrated in Figures \ref{fig5} and \ref{fig6}, the GW signal presents a second peak which strongly depend on the observer angle. Thus, the observer angle can be determined precisely by observing the GW signal. In addition, by observing the associated  multi-wavelength afterglows, the jet structure can be determined. 
Thus, observations of the GW signal may provide us with a precise estimate of the volumetric rate of GRBs. 

The predicted GW signal is below the detection limits of LIGO/VIRGO, KAGRA and similar Earth-based detectors, and is expected to be seen by lower-frequency space-based detectors as BBO and DECIGO. Future detections of GWs from GRBs may provide information on optically thick regions impossible to explore by electromagnetic radiation, clarifying the jet duration, the structure of the progenitor star and the jet acceleration process. 
It is also worth pointing out that the GW detectability can be improved with a network of interferometers. With the rough rule that, the SNR achievable with a network of identical interferometers is the single interferometer SNR multiplied by the square root of the number of interferometers in the network.

\section*{Acknowledgements}

We acknowledge the anonymous referee for a careful reading of the manuscript and for suggestions that improved it substantially.
We acknowledge the computing time granted by DGTIC UNAM on the supercomputer Miztli (project LANCAD-UNAM-DGTIC-281). GU and FDC acknowledge support from the UNAM-PAPIIT grant AG100820 and IG100422. GU acknowledges support from a CONACyT doctoral scholarship. This work was supported by the CONACyT Network Project No. 376127: {\it Sombras, lentes y ondas gravitatorias generadas por objetos compactos astrofísicos}. C.M. thanks PROSNI-UDG support.

\section*{Data availability}

The data underlying this article will be shared on reasonable request to the corresponding author.

\bibliographystyle{mnras}
\bibliography{main} 


\bsp	
\label{lastpage}
\end{document}